\documentclass{article}

\usepackage[]{graphicx}

\newcommand{\be}{\begin{equation}}
\newcommand{\ee}{\end{equation}}

\begin{document}


\title{Magnitude-Redshift Relation for SNe\,Ia, Time Dilation, and Plasma Redshift}         
\author{Ari Brynjolfsson \footnote{Corresponding author: aribrynjolfsson@comcast.net}}

\date{\centering{Applied Radiation Industries, 7 Bridle Path, Wayland, MA 01778, USA}}          

\maketitle

\begin{abstract}    
We have previously shown that the type Ia supernovae data by Riess et al.~match the prediction of the magnitude-redshift relation in the plasma-redshift cosmology.  In this article, we also show that the recent SNLS data, which have a slightly narrower distribution as reported by Astier et al. in 2005, match the predictions of the plasma-redshift cosmology.  The standard deviation of the SNLS-magnitude from the predicted curve is only about 0.14.  The data indicate that there is no cosmic time dilation.  The big-bang cosmology therefore appears false.  The plasma redshift, which follows from exact evaluation of photons interaction with hot sparse electron plasma, leads to a quasi-static, infinite, and everlasting universe.  It does not need big bang, dark energy, or dark matter for describing the observations.  It predicts intrinsic redshifts of galaxies consistent with what is observed.  The Hubble constant that best fits the SNLS data is about 63 km per sec per Mpc.  This corresponds to an average electron density of about 0.0002 per cubic centimeter in intergalactic space.  This density together with the plasma redshift heating to an average plasma temperature in intergalactic space of about 3 million K explains the observed isotropic cosmic microwave background (CMB) and the cosmic X-ray background.

\end{abstract}




\makeatletter	   
\renewcommand{\ps@plain}{
     \renewcommand{\@oddhead}{\textit{Ari Brynjolfsson: Magnitude-redshift relation for SNe\,Ia, time dilation, and plasma redshift}\hfil\textrm{\thepage}}%
     \renewcommand{\@evenhead}{\@oddhead}
     \renewcommand{\@oddfoot}{}
     \renewcommand{\@evenfoot}{\@oddfoot}}
\makeatother     

\pagestyle{plain}


\section{Introduction}

Initially, a Type Ia supernova (SN\,Ia) was considered to be a standard candle.  A closer scrutiny of the nearby supernovae showed however that the maximum luminosity increased significantly with the width (in days) of the light curve.  Many parameters, such as the mass, elemental composition, temperature, strength of the magnetic field, and rotational energy of the progenitor (usually thought to be an accreting white dwarf) can affect the size of the explosion.  The many nuclear reactions resulting in a broad spectrum of reactions products will also have a statistical variance.  The apparent maximum luminosity is also affected by the direction of the magnetic field and the rotational axes relative to the line of sight to the observer.  The larger explosions result in larger widths, $s{\rm{,}}$ of the light curve, and larger maximum luminosities, $ L_{max}(t) = \int L_{max}(t, \, \lambda ) \cdot d \lambda \, {\rm{.}}~$  

\indent  Cosmic time dilation is an integral part of the contemporary big-bang cosmology.  It stretches the light-curve width, $s{\rm{,}}$ by the factor $(1+z) {\rm{.}}~$ For example, Peebles [1] (see pages 91 and 92 of that source) states that the received energy flux $f$ should vary: ``with time as $f \propto L(t/(1+z)) {\rm{.}}~$ That is, the timescale is predicted to be dilated by the redshift, which may be a testable effect.''

\indent  Using nearby supernovae, the researchers seek to obtain $L(t,\, s)$ as a function of time $t$ and the stretch factor $s {\rm{.}}~$ For large $s {\rm{,}}$ however, this is difficult because large $s {\rm{}}$-values are rare.

\indent   If during 15 years we observe 40 SNe\,Ia in the redshift interval from $0.00 \leq z \leq 0.12{\rm{,}}$ then the big-bang hypothesis with its time dilation leads to expectation of about 4000 SNe\,Ia in the same time span and in the redshift interval from $1 \leq z \leq 1.12{\rm{.}}~$ Actually, however, we observe very few.  The few (less than about 0.5\,\%) that we observe are likely to be among the very brightest and therefore with the large stretch factors, $s {\rm{.}}~$ This Malmquist bias must be taken into account.  The probability for observing supernovae SNe\,Ia with a corresponding large stretch factor and luminosity in the interval $0.00 \leq z \leq 0.12{\rm{}}$ is very small, or less than about 1 in 75 years.  This makes it difficult at low redshifts to obtain experimentally a good determination of $L(t,\, s)$ for large $s {\rm{.}}~$

\indent  For nearby supernovae the good analysis of the data by Guy et al.\,[2] (see in particular Figs.\,4 and 6 of that source) shows that the change $(\Delta m - \beta \times c)$ in magnitude versus $s {\rm{}}$ can be approximated by $a'\cdot\log s\approx -2.5\cdot\log s {\rm{.}}~$ In the analyses by Guy et al., the term $-\beta \times c$ corrects for the variation in the color.\,[2]~ For more distant supernovae, for which the light-curve width is $w = s(1+z){\rm{,}}$ we should therefore in the big-bang cosmology expect that the change in magnitude is: $a'\cdot\log s(1+z) \approx -2.5\cdot\log(s) - 2.5 \cdot \log (1+z) {\rm{,}}$ where $s$ should increase with $z {\rm{.}}~$  

\indent  However, Goldhaber et al.\,[3] show in their Fig.\,3 that the light curve width, $w = s (1+z){\rm{,}}$ only doubles when $z$ increases from $z = 0$ to $z = 1 {\rm{.}}~$ If time dilation applies, we must conclude with Goldhaber et al.~that Fig.\,3 in their paper shows that the stretch factor, $s {\rm{,}}$ is a constant as $z $ increases.  However, the constancy of $s$ is contrary to expectations and contrary to the finding by Guy et al.\,[2], see their Figs.\,4 and 6.  We cannot ignore the Malmquist bias, and $s$ must therefore increase with $z {\rm{.}}~$ From the experimental data by Goldhaber et al.~we are inclined to conclude that $w$ is independent of the time dilation, and that it is $s$ that doubles when $z$ increases from $z = 0$ to $z = 1 {\rm{.}}~$ Therefore, their Fig.\,3 indicates, contrary to their conclusion, that there is no cosmic time dilation.  The doubling in the observed width, $w {\rm{,}}$ corresponds then to doubling in $s {\rm{,}}$ which is consistent with the above mentioned analysis by Guy et al. for nearby supernovae.\,[2]~ We conclude therefore that the width is independent of the time dilation factor $(1+z) {\rm{.}}~$ The analyses by Goldhaber et al.\,[3] and Guy et al.\,[2] shows thus that the time dilation is false.

\indent  Foley et al.\,[4] contend that they have "a definitive measurement of time dilation".  They pick a SN\,Ia and find that its features are consistent with time dilation.  This is analogous to picking a white, blond Swede and concluding that all white people are blond.  Even if the features were usually consistent with time dilation, it is possible, but it does not follow that the time dilation is correct. 

\indent  If the quasars are at cosmological distances in accordance with the usual consensus in the astronomical community, then we have, as Hawkins [5] showed, that the time scale of quasar variation does not increase with the redshift as required by the cosmic time dilation.  This good independent study thus contradicts cosmic time dilation, and thus indicates that the time dilation is false.

\indent  Peebles\,[1] (see in particular Eqs.\,(6.41) to (6.44) of that source) shows that according to big-bang cosmology, the surface-intensity integrated over frequencies is $ I = (1+z)^{-4}\,I_0 {\rm{,}}$ where $I_0$ is the corresponding intensity in a non-expanding cosmology, such as in the plasma redshift cosmology.  The factors, $(1+z)^{-1}~{\rm{and}}~(1+z)^{-1} {\rm{,}}$ are due, respectively, to time dilation and to the redshift.  The magnitude-redshift relation for $\Lambda = 1$ and no curvature in the big-bang cosmology is numerically about equal too the corresponding relation in the plasma redshift cosmology, as Eqs.\,(2), (3) and (4) below show.  The remaining factor, which is due to the effect of cosmic time dilation on the observed surface area, reduces the surface intensity by $(1+z)^{-2} {\rm{.}}~$ Recently, Eric Lerner [6] has compared these relations with great many observations and finds that they contradict the prediction of the time dilation.  Thus, also his independent analysis indicates that the cosmic time dilation is false.  

\indent  Therefore, we must correct the absolute magnitude $M_{exp} {\rm{}}$ at maximum brightness, as determined by the supernovae researchers.  These researchers assume time dilation and divide therefore the observed width, $w {\rm{,}}$ of the light curve by the cosmic time dilation factor, $(1+z) {\rm{,}}$ to obtain the stretch factor, $s = w/(1+z) {\rm{.}}~$ From this reduced stretch factor they then determined the absolute magnitude $M_{exp} {\rm{.}}~$ If there is no cosmic time dilation, then both the stretch factor, $s = w {\rm{,}}$ and the maximum brightness estimate would be greater.  We find it reasonable to extrapolate the experimentally determined magnitude-$s$ relation by Guy et al.\,[2] for nearby supernovae (see Fig.\,4 and 6 in [2]) to the larger $s = w $-values that are observed for the more distant supernovae.  Accordingly, the actual absolute magnitude, $M {\rm{,}}$ at maximum intensity of the light curve can be approximated by
\be
M  \approx  M_{exp} - a \log (1+z)  \approx  M_{exp} - 2.5 \log (1+z) \, ,
\ee
where the absolute magnitude $M_{exp} {\rm{}}$ is that estimated by the supernovae researchers.


\section{Redshift-magnitude relation}

\indent  In the plasma-redshift cosmology, the magnitude-redshift relation is given by (see section 5.8 in reference [7]) 
\be
m = 5 \log \left( \frac{{c \cdot 10^{6} }}{{H_0 }} \, {\ln \left( {1 + z} \right)} \right)  + 2.5 \log \left( {1 + z} \right) + 5 \log \left( {1 + z} \right) + M  - 5 \,{\rm{.}} 
\ee
\noindent  On the left side $m = m_{obs} - 1.086 a {\rm{,}}$ where $m_{obs}$ is the observed magnitude and $a{\rm{}}$ the extinction which should include the Compton and Rayleigh scatterings on bound electrons in atoms.  The first term on the right side is due to the distance, $d = (c \cdot 10^{6} /H_0) \ln (1+z)~{\rm{pc}}{\rm{.}}~$  The velocity of light, $c{\rm{,}}$ is in ${\rm{ km}\,\rm{s^{-1}}}$, and the Hubble constant $H_0$ in ${\rm{ km}\,\rm{s^{-1}}\,\rm{Mpc^{-1}}}{\rm{.}}~$ The second term, $2.5 \log (1+z){\rm{,}}$ is due to reduction in photon energy by plasma redshift, and the third term, $5 \log (1+z){\rm{,}}$ is due to removal of photons through Compton scattering on the free electrons in the intergalactic plasma.  $M$ is the absolute magnitude; and the term $-5 = 2.5 \log (1/10^2)$ is caused by definition of $M$ at 10 parsec.

\indent  This magnitude-redshift relation in the plasma-redshift cosmology may be compared with a corresponding big-bang relation for a flat (no curvature) uniformly expanding universe, which corresponds to $\Omega_{\Lambda} = 1$ and $\Omega_{\rho} = 0 {\rm{,}}$ (see Eq.\,(23) of Sandage [8]) 
\be
m = 5 \log \left( {\frac{{c \cdot 10^{6} }}{{ H_0 }}} \, z \right) + 2.5 \log \left( {1 + z} \right) + 2.5 \log \left( {1 + z} \right) + M - 5 {\rm{,}}
\ee
\noindent  where the first term on the right side is due to the distance $d = (c \cdot 10^{6} /H_0)\,z {\rm{.}}~$ The second term is due to the cosmological redshift and the third term is due to the cosmological time dilation.  Supernovae researchers often use the luminosity distance, which is defined as $d_L = d \cdot (1+z){\rm{.}}~$ The 3 first terms on the right side are then replaced by one term.

\indent  When we subtract Eq.\,(2) from Eq.\,(3), we get a very small value, or
\be
\Delta m_d  = 5\log {\frac{{z}}{{\,\ln \left( {1 + z} \right)}} } - 2.5\log \left( {1 + z} \right){\rm{.}}
\ee
\indent  This function is illustrated in Table 1.  


\begin{table}[h]
\centering
{\bf{Table 1}} \, \, The variation in $\Delta m_d$ with the redshift, $z{\rm{,}}$ as defined in Eq.\,(56).

\vspace{2mm}

\begin{tabular}{llllllll}
\hline
$z$ & $\Delta m_d$ & $z$ & $\Delta m_d$ & $z$ & $\Delta m_d$ & $z$ & $\Delta m_d$ \\
\hline \hline
0.1 & 0.0008 & 0.6 & 0.0200 & 1.1 & 0.0496 & 1.6 & 0.0820 \\
0.2 & 0.0030 & 0.7 & 0.0254 & 1.2 & 0.0560 & 1.7 & 0.0885 \\
0.3 & 0.0062 & 0.8 & 0.0312 & 1.3 & 0.0624 & 1.8 & 0.0951 \\
0.4 & 0.0102 & 0.9 & 0.0371 & 1.4 & 0.0689 & 1.9 & 0.1016 \\
0.5 & 0.0149 & 1.0 & 0.0433 & 1.5 & 0.0754 & 2.0 & 0.1081 \\
\hline
\end{tabular}
\end{table}
\noindent  We see that the difference in the observed magnitude, $\Delta m_d {\rm{,}}$ is very small.  That is, the magnitude-redshift relation in plasma redshift is almost identical to that in the big-bang cosmology with $\Omega_{\Lambda} = 1$ and $\Omega_{\rho} = 0 {\rm{.}}~$

\indent  The main difference in the two radically different theories is due to differences in the estimates of the absolute magnitudes.  In plasma redshift theory $M=M_{exp} - 2.5 \cdot \log(1+z){\rm{,}} $ see Eq.\,(1); while in the big-bang cosmology the estimate is equal to $M=M_{exp} {\rm{.}}~$ The difference $(m-M) - (m-M_{exp}){\rm{}}$ is then $+2.5 \cdot \log (1+z) - \Delta m_d {\rm{.}}~$ For z equal to:  0.1. 0,5, 1.0, 1.5, and 2, we get that the difference is: 0.1027, 0.4253, 0.7093, 0.9195, and 1.0847, respectively.


\section{The data from the Supernova Legacy Survey}


\begin{figure}[t]
\centering
\includegraphics[scale=.5]{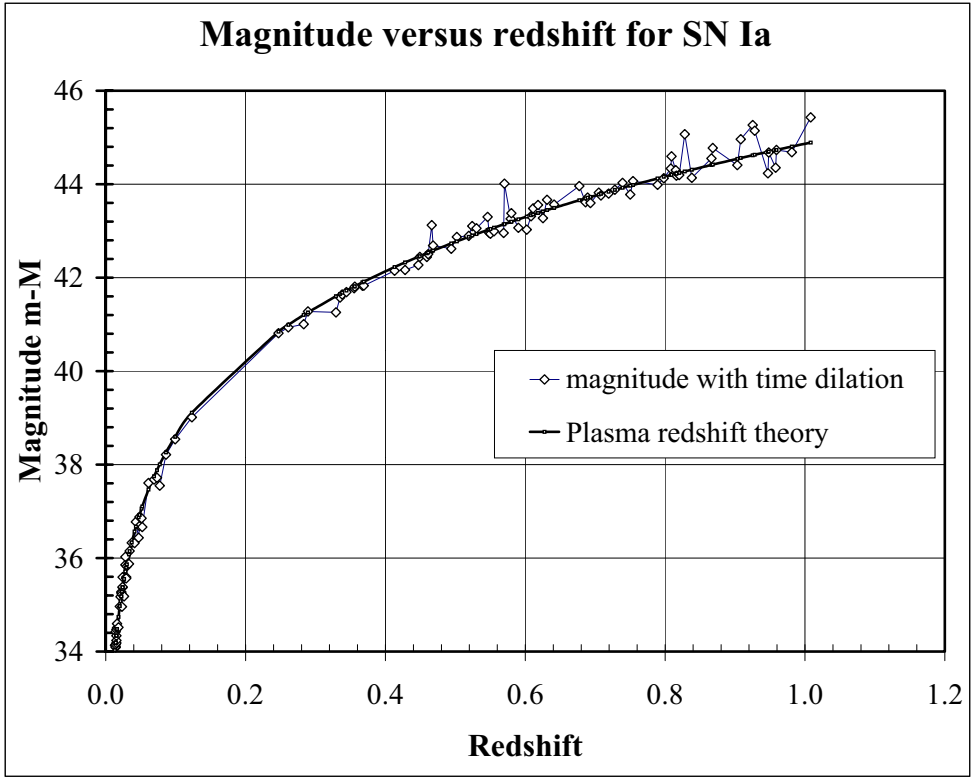}
\caption{The magnitudes, $m-M {\rm{,}}$ of SNe\,Ia on the ordinate versus their redshifts $z'$ on the abscissa.  The data include all 117 (73 high redshift and 44 low redshift) supernovae, as reported by Astier et al.\,[9] (see Tables 8 and 9 of that source).  The diamonds mark the experimental points, while the heavy black curve shows the theoretical predictions of the plasma-redshift theory in accordance with Eq.\,(2).  The Hubble constant, which is determined from the best fit to all 117 supernovae, is ${\rm{H}}_0 = 62.6 ~{\rm{km\,s}}^{-1}\,{\rm{Mpc}}^{-1} {\rm{.}}~$ The standard deviation in a single measurement is $\sigma = 0.23{\rm{.}}$ }
\vspace{2mm}
\end{figure}

\indent   In spite of the small differences between the observed magnitude-redshift relation in plasma-redshift cosmology and contemporary big-bang cosmology, the good quality of the supernovae data makes it possible to distinguish between the two.  

\indent  We have previously perused the data reported in references [10-15] and shown that the data are consistent with no time dilation in the plasma redshift cosmology, see [16,\,17] and in particular section 5.9 and Fig.\,6 in [7].  

\indent  In a recent article Astier et al.\,[9], as a part of the Supernova Legacy Survey (SNLS), selected the most well determined supernovae.  In their table 8 they list 44 nearby SN\,Ia supernovae at redshifts in the range $(0.015\leq z \leq 0.125) {\rm{;}}$ and in their Table 9 they list 73 more distant SN\,Ia supernovae with redshifts in the range $(0.249 \leq z \leq 1.010){\rm{.}}~$ We will compare these well analyzed supernovae data with the theoretical predictions to see if we can distinguish between the two theories.


\begin{figure}[t]
\centering
\includegraphics[scale=.5]{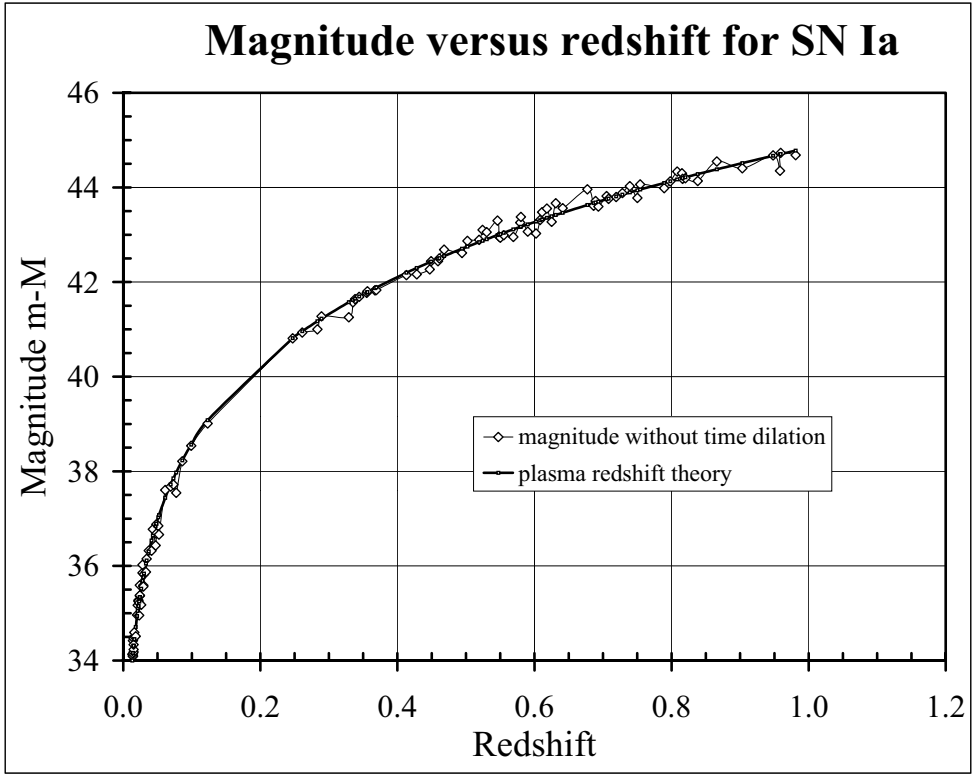}
\caption{The magnitudes, $m-M {\rm{,}}$ versus the redshifts $z'$ on the abscissa.  In this case, we have removed 10 of the high redshift supernovae (shown in Table 2) that deviate more than 2.6 standard deviations from the theoretically expected values.  The diamonds mark the experimental points for the remaining 107 supernovae, while the heavy black curve shows the theoretical predictions given by Eq.\,(2) of the plasma-redshift theory.  The best fit to all 107 SNe\,Ia gives $H_0 =63.2$ and $\sigma= 0.136 {\rm{.}}~$  }
\vspace{0mm}
\end{figure}

\indent  To the observer, the inhomogeneity in the density of the plasma along the light path from the supernova causes computational complications.  In the first approximation we consider two groups: the intrinsic redshifts and the average intergalactic or cosmological redshifts.  An intrinsic redshift is caused by the hot and relatively dense galactic corona with electron densities in the range of about $ 10^{-3} \leq N_e \leq 1~{\rm{cm}}^{-3} {\rm{;}}$ while the intergalactic redshifts are caused by hot intergalactic plasma with electron densities usually in the range of about $10^{-5} \leq N_e \leq 10^{-3}~{\rm{cm}}^{-3} {\rm{,}}$ and with an average electron density of about $N_e \approx 2\cdot10^{-4}~{\rm{cm}}^{-3} {\rm{.}}~$ The sum of the intrinsic redshift of the Milky Way and that of the host galaxy for each supernova is assumed to be about $z \approx 0.00185$ provided the Galactic latitude of the supernova exceeds about $b = 20^o {\rm{.}}~$ This limit depends on the longitude.
 
\indent  The intrinsic redshift affects the distance term in Eq.\,(2) but not the other terms.  For estimating the distance term in Eq.\,(2), we use therefore
\be
z' = z_{exp} - 0.00185 .
\ee
\noindent  For the other terms in Eq.\,(2), we use $ z = z_{exp} {\rm{,}}$ where $z_{exp}$ is the value reported by Astier et al.\,[9].

\indent  In Figs.\,1, 2 and 3, we set the absolute magnitude $M = M_{exp} - 2.5 \log (1+z) {\rm{,}}$ where $M_{exp}$ is the absolute magnitude, as listed by Astier et al. in their Tables 8 and 9 of [9].  This about corrects the magnitude for the false time dilation.  For determining the distance, $d {\rm{,}}$ we use $z'$ given by Eq.\,(5) on the abscissa.  But in Fig.\,4, we set $M = M_{exp} $ and $z = z_{exp}{\rm{}}$ on the abscissa.  


\begin{figure}[h]
\centering
\includegraphics[scale=.45]{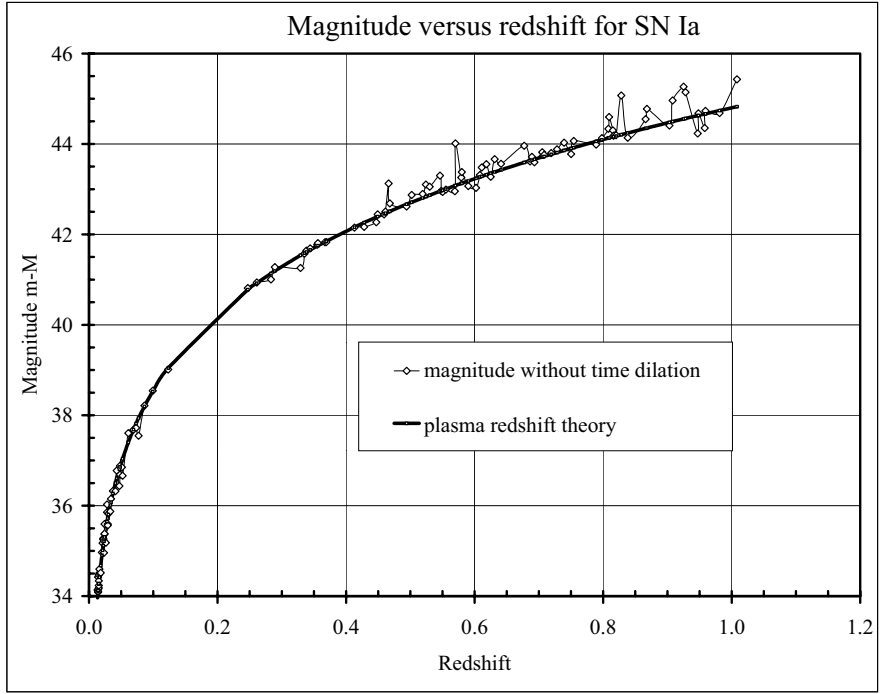}
\caption{The magnitudes, $m-M {\rm{,}}$ versus the redshifts $z'{\rm{.}}~$ The data include all 117 (73 high redshift and 44 low redshift) supernovae, as reported in Tables 8 and 9 by Astier et al.\,[9].  In this case, the Hubble constant, ${\rm{H}}_0 = 64.5 ~{\rm{km\,s}}^{-1}\,{\rm{Mpc}}^{-1} {\rm{,}}$ is derived from best fit to only the 44 nearest supernovae (not all 117 as in Fig.\,1).  The standard deviation in a single measurement is $\sigma = 0.24{\rm{.}}~$ }
\vspace{0mm}
\end{figure}


\begin{table}[h]
\centering
{\bf{Table 2}} \, \, The 10 supernovae with magnitudes that deviate the most from the theoretical curve.

\vspace{2mm}

\begin{tabular}{lllllllll}
\hline
$~~~{\rm{Name}}$& ~~$z$ &$\delta m $& ~~${\rm{Name}}$ &~~ $z$ & $\delta m$ & ~~~${\rm{Name}}$ &~~ $z$ & $\delta m$ \\
\hline \hline
SNLS-03D4au & 0.468 & 0.56     &~SNLS-03D1cm & 0.870 & 0.36  &~SNLS-03D4cx & 0.947 & -0.46   \\
SNLS-03D4bc & 0.572 & 0.87     &~SNLS-04D3gx & 0.910 & 0.40 &~SNLS-04D3dd & 1.010 & 0.54    \\
SNLS-04D4dm & 0.811 & 0.39     &~SNLS-03D4cy & 0.927 & 0.65   & & & \\
SNLS-04D3cp & 0.830 & 0.80     &~SNLS-04D3ki & 0.930 & 0.51   & & & \\   
\hline
\end{tabular}
\end{table}  
 

\begin{figure}[h]
\centering
\includegraphics[scale=.45]{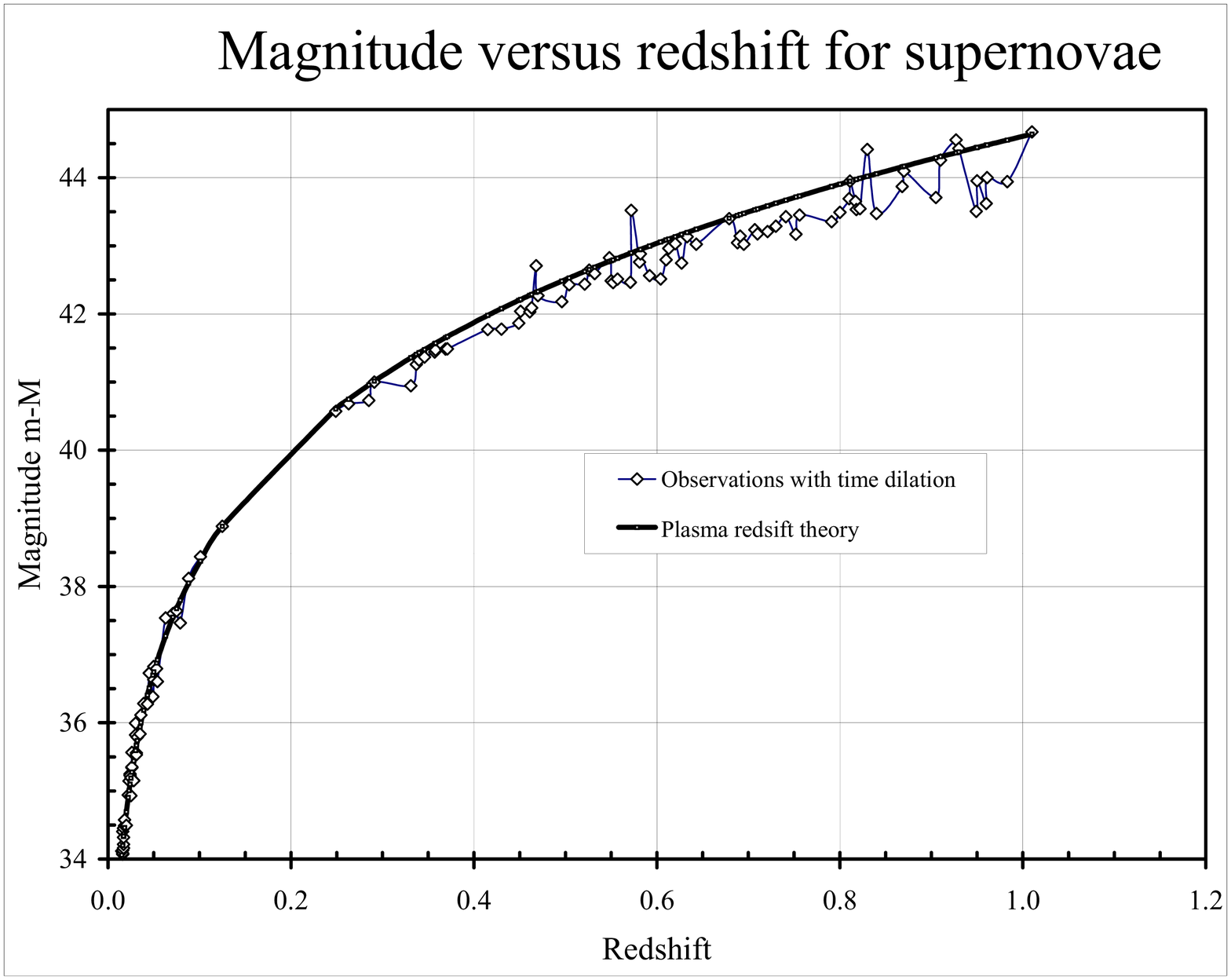}
\caption{The magnitudes, $m-M = m - M_{exp}{\rm{,}}$ versus the redshifts $z=z_{exp}$ on the abscissa.  The data include all 117 (73 high and 44 low redshift) SNe\,Ia, as reported by Astier et al.\,[9].  The Hubble constant, ${\rm{H}}_0 = 70.4 ~{\rm{km\,s}}^{-1}\,{\rm{Mpc}}^{-1} {\rm{,}}$ is derived from best fit to only the 44 nearest supernovae.  For the 6 highest redshift supernovae, the standard deviation in a single measurement is $\sigma = 0.70 {\rm{}}$ in agreement with Eq.\,(1). }
\vspace{0mm}
\end{figure}


\indent  In Fig.\,1 we include all of the 117 supernovae in our evaluation.  We find then that one standard deviation in $m-M {\rm{}}$ is $\sigma = 0.23 {\rm{.}}~$  The 10 supernovae that deviate the most are shown in Table 2.  If we disregard these 10 supernovae, the standard deviation decreases from $\sigma = 0.23$ to $\sigma = 0.136 {\rm{.}}~$  That many supernovae outside the $\delta m = 0.36 $-limit show that the distribution is not Gaussian.  For larger $z$-values the uncertainty in the absolute magnitude increases.  The 9 positive values versus 1 negative value also indicate that the distribution is skewed.  Most likely the absorption in the supernova host galaxy is not adequately taken into account, especially, for SN\,Ia at large distances.  These SN\,Ia are at high Galactic latitudes.  The extra absorption (if any) could be due to the supernova being beyond the center of the host galaxy.  The Compton and Rayleigh scatterings on bound electrons in a low-density gas in the host galaxy are nearly colorblind.  It is therefore difficult to take these absorptions into account, but they could be significant, especially, if the supernova is in the back of the host galaxy.  In their analyses, Astier et al. usually disregarded the supernovae SNLS-03D4au and SNLS-03D4bc, which deviate by 0.56 and 0.87 magnitudes, respectively, from the fitted curve.

\indent  Astier et al.\,[9] used a value ${\rm{H}}_0 = 70~{\rm{km\,s}}^{-1}\,{\rm{Mpc}}^{-1} {\rm{}}$  for the Hubble constant.  In plasma redshift theory, the Hubble constant is partly reduced by the correction of the absolute magnitude, and partly by the intrinsic redshift.  When the supernovae with the largest deviation are eliminated, as in Fig.\,2, the standard deviation in magnitude $m-M$ is reduced from $\sigma = 0.23 {\rm{,}}$ to about $\sigma = 0.136 {\rm{,}}$ magnitudes with the 10 of the 117 supernovae deviating more than 2.6 standard deviations.  The removal of these 10 supernovae increases the Hubble constant from 62.6 to about 63.2.

\indent  In Fig.\,3, we plot the magnitude $m-M$ corrected for the time dilation of all the 117 supernovae versus their redshift, $z'=z-0.00185 {\rm{.}}~$  $H_0 = 64.5 {\rm{}}$ is optimized for the best fit to only the 44 low-redshift supernovae.  The standard deviation for a single supernovae is $\sigma = 0.24{\rm{.}}~$ 

\indent In Fig.\,(4), we plot similarly $m-M = m-M_{exp}$ uncorrected for the time dilation of all the 117 supernovae versus their redshift, $z=z_{exp} {\rm{.}}~$  $H_0 = 70.4 {\rm{}}$ is optimized for the best fit to only the 44 low-redshift supernovae as in Fig.\,3.  The averaged standard deviation for the 6 highest redshift supernovae is $\sigma = 0.70 {\rm{.}}~$  This is consistent with Eq.\,(1).  Figs.\,3 and 4 therefore indicate that the cosmic time dilation is false.


\section{Conclusions and discussions}

\indent  The very best data by the supernova researchers are consistent with the magnitude-redshift relations predicted by the plasma redshift.  The data indicate that there is no time dilation; that is, the data indicate that the contemporary big-bang hypothesis is false. 

\indent  In Figs.\,1, 2, and 3 it is assumed that each galaxy has an intrinsic redshift of about $z = 0.000925{\rm{,}}$ which was derived independently from the density determination in the Galactic corona.\,[7]~ Fig.\,1 to 3 are consistent with these intrinsic redshift estimates.  Fig.\,4 indicates that Eq.\,(1), which eliminates the time dilation from the magnitude determination, is a good approximation.

\indent   The 10 high-redshift supernovae with excessive deviation from the theoretical curve are listed in Table 2.  These 10 supernovae are all at high Galactic latitudes, 9 have positive and 1 negative deviations.  This suggests that a large positive deviation is due to an underestimate of the absorption in the neutral gas of host galaxy.  Fig.\,2 shows that when we exclude these supernovae, both the low and high-redshift supernovae are close to the theoretical curve for plasma redshift.  The Hubble constant derived from the best fit to the remaining 107 supernovae is ${\rm{H}}_0 = 63.2 ~{\rm{km\,s}}^{-1}\,{\rm{Mpc}}^{-1} {\rm{.}}~$ The overall standard deviation in the magnitude $m-M$ of a single SN\,Ia is only $\sigma = 0.14 {\rm{.}}~$

\indent  Using exact calculations, plasma redshift follows from conventional axioms of physics.\,[7]~ From $H_0 = 63.2 {\rm{,}}$ we derive an average electron density of $(N_e)_{av}=H_0/(3.076 \cdot 10^5) = 2.05\cdot 10^{-4}~{\rm{cm}}^{-3} {\rm{}}$ in intergalactic space (see Eq.\,(49) in [7]).  The energy loss of photons in the plasma redshift is absorbed in the plasma.  The corresponding heating leads to an average plasma temperature of about 3 million K in intergalactic space.\,[7]~ These densities and temperatures of the intergalactic plasma explain the isotropic CMB and the X-ray background, as shown in sections 5.10 and 5.11 in reference [7]. The plasma redshift cosmology thus gives a coherent prediction of the observations.


\end{document}